\documentclass{ws-ijqi}

\def\be{\begin{equation}}
\def\ee{\end{equation}}

\begin{document}

\markboth{Pasquale Calabrese}
{Entanglement Entropy}

\catchline{}{}{}{}{}

\title{Entanglement Entropy and Quantum Field Theory:\\ A Non-Technical Introduction}

\author{Pasquale Calabrese}
\address{Rudolf Peierls Centre for Theoretical Physics,
         1 Keble Road, Oxford OX1 3NP, UK\\
calabres@thphys.ox.ac.uk}

\author{John Cardy}
\address{Rudolf Peierls Centre for Theoretical Physics,
         1 Keble Road, Oxford OX1 3NP, UK and All Souls College, Oxford}

\maketitle


\begin{abstract}
In these proceedings we give a pedagogical and non-technical 
introduction to the Quantum Field Theory approach to entanglement
entropy.
Particular attention is devoted to the one space dimensional 
case, with a linear dispersion relation, that, at a quantum critical
point, can be effectively described by a two-dimensional Conformal 
Field Theory. 
\end{abstract}


\section{Introduction}

There has been considerable interest in formulating measures of
entanglement and applying them to extended quantum systems with
many degrees of freedom, such as quantum spin chains, especially close to 
a quantum phase transition (see, e.g., several other reports in these 
proceedings). 
One of these measures is entanglement entropy.\cite{Bennett} 
Suppose the whole system is in a pure quantum state $|\Psi\rangle$, with 
density matrix $\rho=|\Psi\rangle\langle\Psi|$, and an observer A measures 
a subset $A$ of a complete set of commuting observables, while another
observer B may measure the remainder. A's reduced density matrix is
$\rho_A={\rm Tr}_B\,\rho$. The entanglement entropy is just the von
Neumann entropy $S_A=-{\rm Tr}_A\,\rho_A\log\rho_A$ associated with this
reduced density matrix.

Although there are other measures of entanglement,
the entropy is most readily suited to analytic investigation.
Lack of space prevents us from discussing other measures of entanglement 
and for referring properly to all the works about entanglement entropy.
For a more complete list of references we refer, e.g., to our 
papers.\cite{cc-04,cc-05}

In these proceedings we try to give a pedagogical and non-technical 
introduction to the Quantum Field Theory (QFT) approach to entanglement
entropy.\cite{cc-04} 
Particular attention is devoted to the one space dimensional 
case, with a linear dispersion relation, that, exactly at the quantum phase 
transition, can be effectively described by a two-dimensional Conformal 
Field Theory (CFT).
Again, for lack of space, we will not discuss the interesting topic of 
the time evolution of entanglement entropy\cite{cc-05} and we will limit
ourselves to discussion of the ground state properties.

The layout of the paper is as follows. In the next section, we 
give an introduction to the concepts of renormalization group 
and conformal invariance. 
In section \ref{pathsec} we
give a path integral formula for the entanglement entropy. 
In Section \ref{cftsec}, we apply this formula to the calculation of the 
entanglement entropy in 1+1 dimensional CFT. In section \ref{noncrit}
we relax the conformal constraint and study the more general case of a 
system close, but not exactly at the phase transition.
Several omitted (technical) details, as well as some results in higher 
dimensions, can be found in the original paper.\cite{cc-04}

\section{Renormalization Group and Conformal Transformations}

For someone who is not familiar with the modern theory of phase transitions,
the first natural question arising after reading the Introduction is 
what is the connection between the QFT and a 
many-body system undergoing a {\it continuous} phase transition. 
The bridge between these apparently disconnected subjects is given by
the Renormalization Group (RG) theory.\cite{cardybook}
Let us consider a quantum model defined by an hamiltonian $H(g)$, where
$g$ is a tunable experimental parameter, e.g. in the well-known Ising model
in a transverse field it is the magnetic field in the transverse 
direction.\cite{sach}
Such hamiltonian at $g=g_c$ undergoes a continuous phase transition 
($g_c$ is called Quantum Critical Point, QCP).
Close to the QCP, the correlation length, 
that is the only relevant scale for the long-distance physics, 
behaves like $\xi\sim |g-g_c|^{-\nu}$, diverging at the QCP. 
($\nu>0$ is an example of {\it critical exponent}.)
Thus, at the QCP, the system is {\it scale invariant}. 
Now, the universality hypothesis states that some physical properties (called
universal), close to the phase transition, do not depend on 
microscopic details, but only 
on global properties, such as symmetries and dimensionality.
This hypothesis has an elegant explanation in terms of RG theory:
Under RG transformations (that, roughly speaking, are practical 
implementations of scale transformations), 
different hamiltonians sharing the same universal 
characteristics flow to the same {\em fixed point}. 
This completely determines the long-distance behavior. 
For example a model defined on the lattice is
defined by an hamiltonian which is invariant under translations multiple of 
the lattice spacing. The resulting fixed point hamiltonian, instead is 
generically invariant under arbitrary translations, allowing for the use of 
a continuum field theory.
For the same reason, the critical point hamiltonian is usually invariant 
under general rotations.

These transformations of rotations, translations, and scaling form a 
group. Let us exploit the consequences of this symmetry group.
Consider the two-point function of a scalar observable 
$\langle \phi({\bf r}_1) \phi({\bf r}_2) \rangle$.
By translational invariance it can be only a function of 
${\bf r}_1- {\bf r}_2$, by rotational invariance can depend only upon
the modulus of such vector, and for a scale transformation 
${\bf r}\to b{\bf r}$ it must behaves like
\be
\langle \phi({\bf r}_1) \phi({\bf r}_2)\rangle=b^{2 \Delta_\phi}
\langle \phi(b {\bf r}_1) \phi(b{\bf r}_2)\rangle\,,
\label{scaltra}
\ee
where the exponent $\Delta_\phi$ is called scaling dimension of the field $\phi$.
These three conditions can be true if and only if
\be
\langle \phi({\bf r}_1) \phi({\bf r}_2)\rangle=
|{\bf r}_1-{\bf r}_2|^{-2 \Delta_\phi}\,,
\label{2ptsca}
\ee
apart a normalization constant we set equal to $1$.

It turns out that a fixed point hamiltonian that is invariant under
translations, rotations, and scaling transformations has usually the symmetry 
of the larger {\it conformal} group\cite{confbook} defined as the set of 
transformations that do not change the angles between two arbitrary curves 
crossing each other in some point.
The consequences under this further invariance in a two-dimensional 
euclidean space\footnote{In a one-dimensional quantum system with a linear
dispersion relation $E\propto k$ (this to ensure that space and time scale 
in the same manner), the two euclidean coordinates $(x,y)$ correspond to the 
space and to the imaginary time $\tau$, obtained by Wick rotation of the real
 time $t$, i.e. $y=\tau\equiv i t$.}
 are extraordinary. 
In fact, in two-dimensions, we can use complex coordinates $z=x+iy$ and
$\bar{z}=x-iy$.
It is straightforward to prove that all the analytic functions $f(z)$ 
are conformal transformations (this is the reason such mapping are so useful 
to solve Laplace's equations).
The resulting symmetry group is infinite dimensional, and we can calculate, 
in principle, everything in an analytic way. 
For example, under a transformation of the type $z\to w=w(z)$, 
we can generalize Eq. (\ref{scaltra}) to a space dependent 
scale factor $b(z)=w'(z)$ (the prime here and in the following denotes the
derivative with respect to $z$), obtaining
\be
\langle \phi(z_1) \phi(z_2) \rangle=
(|w'(z_1)w'(z_2)|)^{2 \Delta_\phi} 
\langle \phi(w(z_1)) \phi(w(z_2))\rangle\,.
\label{2pt}
\ee
This equation relates the two-point function of a scalar field
on the plane given by Eq. (\ref{2ptsca}) to 
the one in any other geometry (cylinder, torus, \dots).

A crucial rule in CFT is played by the so called stress tensor $T^{\mu\nu}$, 
that can be defined as follows. 
Under an arbitrary transformation $x^\mu\to x^\mu+\epsilon^\mu$,
the euclidean action (the hamiltonian for classical systems) changes as
\be
S\to S+\delta S, \qquad {\rm with }\quad \delta S=
\int d^2x T^{\mu\nu} \partial_\mu\epsilon_\nu\,.
\ee
In general for a CFT $T^{\mu\nu}$ can be chosen to be symmetric and 
traceless.\footnote{Notice that in complex coordinates the traceless condition
reads $T_{z\bar z}=0$. The standard convention is $T(z)=-2\pi T_{zz}$ and
$\overline{T}(\bar{z})=-2\pi T_{\bar{z}\bar{z}}$, called holomorphic 
and antiholomorphic stress tensor respectively.\cite{confbook}}

One of the most intriguing results\cite{confbook} of two-dimensional CFT is 
that the universality class (of minimal unitary models) 
is characterized by a single quantity called 
central charge, that assumes only the discrete values
\be
c=1-\frac{6}{m(m+1)}, \qquad {\rm with} \; m=3,4,\dots,\infty\,.
\label{cunit}
\ee
For example the Ising universality class corresponds to $c=1/2$, the 
free boson to $c=1$, and the three state Potts model to $c=4/5$.
Cases with $c$ different from the value allowed by Eq. (\ref{cunit})
are also of physical interest (e.g. the case $c=0$ is the critical 
percolation), but they are always pathological in some sense.
From the knowledge of the central change $c$ (that actually appears directly 
in several important physical quantities, as the entanglement 
entropy\cite{Holzhey}), we 
can (in principle) determine all the critical properties of the model, such as
the critical exponents.

\section{Path integral formula for the Entanglement Entropy}
\label{pathsec}

Consider a lattice quantum theory in one space and one time
dimension. The lattice spacing is $a$,
and the lattice sites are labelled by a discrete variable $x$. 
The domain of $x$ can be finite, i.e. some interval of length $L$, 
semi-infinite, or infinite. Time is considered to be continuous. 
A complete set of local
commuting observables will be denoted by $\{\hat\phi(x)\}$, and their
eigenvalues and corresponding eigenstates by $\{\phi(x)\}$ and
$\otimes_x|\{\phi(x)\}\rangle$ respectively. For a bosonic lattice
field theory,
these will be the fundamental bosonic fields of the theory; for a spin
model some particular component of the local spin. The dynamics of the
theory is described by a time-evolution operator $\hat H$. The density
matrix $\rho$ in a thermal state at inverse temperature $\beta$
is 
\begin{equation}
\rho(\{\phi(x'')''\}|\{\phi(x')'\})=
Z(\beta)^{-1}\langle\{\phi(x'')''\}|e^{-\beta\hat
H}|\{\phi(x')'\}\rangle\,,
\end{equation}
where $Z(\beta)={\rm Tr}\,e^{-\beta\hat H}$ is the partition function.

This may be expressed in the standard way
as a (euclidean) path integral:
\begin{equation}
\label{pathi}
\rho=Z^{-1}\int[d\phi(x,\tau)]
\prod_x\delta(\phi(x,0)-\phi(x')')\prod_x
\delta(\phi(x,\beta)-\phi(x'')'')\,e^{-S_E}\,,
\end{equation}
where $S_E=\int_0^\beta L_Ed\tau$, with $L_E$ the euclidean lagrangian.

The normalization factor of the partition function ensures that
${\rm Tr}\,\rho=1$, and is found by 
setting $\{\phi(x)''\}=\{\phi(x)'\}$ and integrating
over these variables. This has the effect of sewing together the edges
along $\tau=0$ and $\tau=\beta$ to form a cylinder of circumference
$\beta$.

\begin{figure}[b]
\caption{Path integral representation of $\rho_A$.}
\vspace{2mm}
\centerline{\epsfig{width=9cm,file=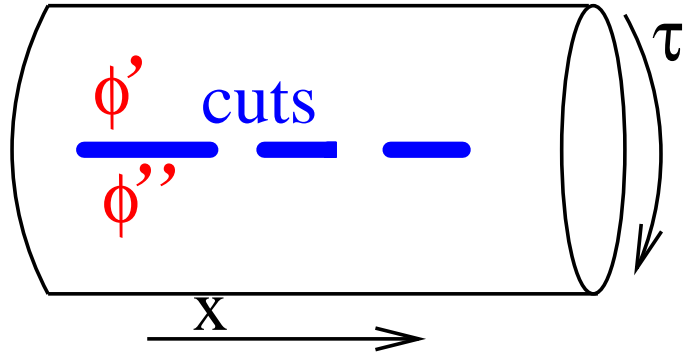}}
\label{cil}
\end{figure}

Let $A$ be a subsystem consisting of the points $x$ in the
disjoint intervals $(u_1,v_1),\ldots,(u_N,v_N)$. An expression for the
the reduced density matrix $\rho_A$
may be found 
by sewing together only those points $x$ which are not in $A$. This
will leave open cuts, one for each interval
$(u_j,v_j)$, along the the line $\tau=0$. See figure \ref{cil} for a 
pictorial representation of this.

We may then compute ${\rm Tr}\,\rho_A^n$, for any positive
integer $n$, by making $n$ copies of the above, labelled by an integer
$k$ with $1\leq k\leq n$, and sewing them together cyclically along the 
the cuts so that $\phi(x)'_k=\phi(x)''_{k+1}$ (and
$\phi(x)'_n=\phi(x)''_1$) for all $x\in A$. Let us denote the
path integral on this $n$-sheeted structure by $Z_n(A)$.
Then
\begin{equation}
\label{ZoverZ}
{\rm Tr}\,\rho_A^n={Z_n(A)\over Z^n}\,.
\end{equation}
Since ${\rm Tr}\,\rho_A^n=\sum_j\lambda_j^n$,
where $\{\lambda_j\}$ are all the eigenvalues of $\rho_A$ (which lie in 
$[0,1)$), it follows that the left hand side is absolutely convergent and 
therefore analytic for all ${\rm Re}\,n>1$. 
The derivative with respect to $n$ therefore also
exists and is analytic in the region. Moreover, if the entropy
$\rho_A=-\sum_\lambda\lambda\log\lambda$ is finite, the limit as
$n\to1^+$ of the first derivative converges to this value. 
We conclude that the right hand side of (\ref{ZoverZ}) has a unique
analytic continuation to ${\rm Re}\,n>1$ and that its first derivative
at $n=1$ gives the required entropy:
\begin{equation}
S_A= -\lim_{n\to1}{\partial\over\partial n} {\rm Tr}\,\rho_A^n=
-\lim_{n\to1}{\partial\over\partial n}{Z_n(A)\over Z^n}\,.
\end{equation}

So far, everything has been in the discrete space domain. 
We showed\cite{cc-04} that the continuum limit can be taken safely, 
since most of the UV divergences of the QFT cancel in the ratio 
(\ref{ZoverZ}).

\section{Entanglement entropy in 1+1-dimensional CFT.}
\label{cftsec}

Now specialize the discussion of the previous section
to the case when the field theory is relativistic and massless, i.e. a
CFT, with central charge $c$.

\subsection{Single interval}

We first consider the case of a single interval of length $\ell$ in an 
infinitely long 1d quantum system, at zero temperature. 
The ratio (\ref{ZoverZ}) is given by $\langle 0|0\rangle_{{\cal R}_n}$, i.e.
the vacuum expectation value in the $n$-sheeted surface. 
Within CFT to obtain this expectation value is enough to know how 
it transforms under a general conformal transformation. 
This is formally given by $\langle T(w) \rangle_{{\cal R}_n}$, where
$T(w)$ is the (holomorphic) stress tensor.
The reason for that will be clear (we hope) in the following.  

To obtain $\langle T(w) \rangle_{{\cal R}_n}$, we need to map the 
$n$-sheeted surface onto a geometry where the mean value of the 
stress tensor is known, and then use the transformation law\cite{confbook} 
\begin{equation}
\label{schwartz}
T(w)=(z'')^2\,T(z)+\frac{c}{12} \frac{z'''z'-\frac32{z''}^2}{{z'}^2}\,.
\end{equation} 

The mapping we need is $w\to z(w)=\big((w-u)/(w-v)\big)^{1/n}$.
This maps the whole of the $n$-sheeted surface ${\cal R}_n$ to the 
$z$-plane $\bf C$, where by translational and rotational invariance 
$\langle T(z)\rangle_{\bf C}=0$.
In fact, the conformal transformation $w\to\zeta=(w-u)/(w-v)$ maps the branch 
points to $(0,\infty)$. This is then uniformised by the mapping 
$\zeta\to z=\zeta^{1/n}$. 

Thus, taking the expectation value of (\ref{schwartz}) and calculating the 
three derivatives, we find
\be
\langle T(w)\rangle_{{\cal R}_n}=
{c(1-(1/n)^2)\over 24}{(v-u)^2\over(w-u)^2(w-v)^2}\,.
\ee
Now the way of working should be clear.
Compare this with the standard form of the correlator of
$T$ with two (primary) operators $\Phi_n(u)$ and $\Phi_{-n}(v)$ which have
the same complex scaling dimensions 
$\Delta_n=\overline\Delta_n=(c/24)(1-(1/n)^2)$ ({\it conformal
Ward identity}):\cite{confbook}
\be
\label{TPP}
\langle T(w)\Phi_n(u)\Phi_{-n}(v)\rangle_{\bf C}=
{\Delta_n\over (w-u)^2(w-v)^2(v-u)^{2\Delta_n-2}
(\bar v-\bar u)^{2\Delta_n}}\,,
\ee
where $\Phi_{\pm n}$ are normalized so that
$\langle \Phi_n(u)\Phi_{-n}(v)\rangle_{\bf C}=|v-u|^{-4\Delta_n}$. 
In writing the above, we are assuming that $w$ is a complex coordinate
on a single sheet $\bf C$, which is now decoupled from the others.   
We have therefore shown that 
\begin{equation}
\langle T(w)\rangle_{{\cal R}_n}\equiv
{\int[d\phi]T(w)e^{-S_E({\cal R}_n)}\over
\int[d\phi]e^{-S_E({\cal R}_n)}}=
{\langle T(w)\Phi_n(u)\Phi_{-n}(v)\rangle_{\bf C}
\over \langle \Phi_n(u)\Phi_{-n}(v)\rangle_{\bf C}}\,.
\end{equation}

The insertion of $T(w)$ on each sheet is given by (\ref{TPP}).
Since this is to be inserted on all the sheets, the right hand side 
gets multiplied by 
a factor $n$.
Since the Ward identity (\ref{TPP}) determines all the properties under 
conformal transformations, we conclude that 
$Z_n({A})/Z^n\propto{\rm Tr}\,\rho_{A}^n$ behaves 
(apart from a possible overall constant)
under scale and conformal transformations identically to the $n$th power
of two-point function of a primary operator $\Phi_n$ with 
$\Delta_n=\overline\Delta_n=(c/24)(1-(1/n)^2)$. 
In particular, this means that
\be
{\rm Tr}\,\rho_{A}^n=c_n\left(\frac{v-u}{a}\right)^{-(c/6)(n-1/n)}\,,
\ee
where the exponent is just $4n\Delta_n$. 
The power of $a$ has been inserted so as the make the final result 
dimensionless, as it should be. 
The constants $c_n$ are not determined by this method. However $c_1$
must be unity. Differentiating with respect to $n$ and setting $n=1$, we
recover the result of Holzhey et al.\cite{Holzhey}
\be
S_A=\frac{c}{3}\log\frac{\ell}{a}+c'_1\,.
\ee
Notice that the constant $c'_1$ is {\it not} universal.

The fact that ${\rm Tr}\,\rho_{A}^n$ transforms under conformal 
transformations as a 2-point function of primary operators $\Phi_{\pm n}$
means that it can be simply computed in other geometries, obtained by a
conformal mapping $z\to w=w(z)$, using equation (\ref{2pt}).

\subsection{Finite Temperature}

The transformation $w\to w'=(\beta/2\pi)\log w$ maps each sheet in
the $w$-plane into an infinitely long cylinder of circumference $\beta$. 
The sheets are now sewn together along a branch cut joining the images
of the points $u$ and $v$. By arranging this to lie parallel to the axis
of the cylinder, we get an expression for ${\rm Tr}\,\rho_{A}^n$
in a thermal mixed state at finite temperature $\beta^{-1}$. 
This leads to the result for the entropy
\begin{equation}
S_A(\beta)\sim
\frac{c}{3}\log\left(\frac{\beta}{\pi a}\sinh\frac{\pi\ell}{\beta}\right)+c_1'\,.
\label{Sbeta}
\end{equation}
For $\ell\ll\beta$ we find $S_A\sim(c/3)\log(\ell/a)$ as before, while, in the
opposite limit $\ell\gg\beta$, $S_A\sim(\pi c/3)(\ell/\beta)$. In this
limit, the von Neumann entropy is extensive, and its density agrees with
that of the Gibbs entropy of an isolated system of length $\ell$.

Notice that the entropy given by Eq. (\ref{Sbeta}) is a measure of quantum
entanglement only for $\beta\to\infty$. In the opposite limit $\beta\ll\ell$
it is just a measure of the classical entropy. Eq. (\ref{Sbeta})
tells us how the crossover between these two different objects is realized
increasing the temperature.

\subsection{Finite Systems}

Orienting the branch cut perpendicular to the axis
of the cylinder (with the replacement $\beta\to L$) corresponds
to the entropy of a subsystem of length $\ell$ in a finite system of
length $L$, with periodic boundary conditions. 
This gives 
\begin{equation}
S_A\sim \frac{c}{3}\log\left(\frac{L}{\pi a}\sin\frac{\pi\ell}{L}\right)+c_1'\,.
\end{equation}
Note that this expression is symmetric under $\ell\to L-\ell$. It is
maximal when $\ell=L/2$. 

\subsection{Finite system with a boundary.}

Next consider the case when the 1d system is a semi-infinite line,
say $[0,\infty)$, and the subsystem $A$ is the finite interval
$[0,\ell)$. 
At $x=0$ a boundary condition is imposed, that must be conformal invariant 
as well.
The $n$-sheeted surface then consists of $n$ copies
of the half-plane $x\geq0$, sewn together along $0\leq x<\ell, \tau=0$.
Once again, we work initially at zero temperature. It is convenient
to use the complex variable $w=\tau+ix$. The uniformising transformation
is now $z=\big((w-i\ell)/(w+i\ell)\big)^{1/n}$, which maps the whole
$n$-sheeted surface to the unit disc $|z|\leq1$. In this geometry, 
$\langle T(z)\rangle=0$ by rotational invariance, so that, using 
(\ref{schwartz}), we find
\begin{equation}
\label{hp}
\langle T(w)\rangle_{{\cal R}_n}=
{\Delta_n(2\ell)^2\over(w-i\ell)^2(w+i\ell)^2}\,,
\end{equation}
where $\Delta_n=(c/24)(1-n^{-2})$ as before. 
(\ref{hp}) has the same form as $\langle T(w)\Phi_n(i\ell)\rangle$,
which follows from the Ward identities of boundary CFT,\cite{confbook} 
with the normalization $\langle\Phi_n(i\ell)\rangle=(2\ell)^{-\Delta_n}$. 

The analysis then proceeds in analogy with the previous case. We find
\begin{equation}
{\rm Tr}\,\rho_{A}^n\sim \tilde c_n(2\ell/a)^{(c/12)(n-1/n)}\,,
\end{equation}
so that
$S_A\sim(c/6)\log(2\ell/a)+{\tilde c}'_1$. 

Once again, this result can be conformally transformed into a number of
other cases. At finite temperature $\beta^{-1}$ we find
\begin{equation}
S_A(\beta)\sim(c/6)\log\big((\beta/\pi a)\sinh(2\pi\ell/\beta)\big)+
{\tilde c}'_1\,.
\end{equation}
By taking the limit when $\ell\gg\beta$ we find the same extensive
entropy as before. 

For a completely finite 1d system, of length $L$ (with two conformal boundary 
conditions at $x=0$ and $x=L$), at zero temperature, divided into two pieces 
of lengths $\ell$ and $L-\ell$, we similarly find
\begin{equation}
S_A=(c/6)\log\big((2L/\pi a)\sin(\pi\ell/L)\big)+{\tilde c}'_1\,.
\end{equation}

\subsection{General case}

For the general case, when $A$ consists of several disjoint intervals 
$(u_k,v_k)$, the uniformising transformation is rather 
complicated,\cite{cc-04} but the method is the same. 
After a quite long algebra we found
\be
\label{general}
S_A=\frac c3\left(\sum_{j\leq k}\log\frac{v_k-u_j}{a}
-\sum_{j<k}\log\frac{u_k-u_j}{a}-\sum_{j<k}\log\frac{v_k-v_j}{a}
\right)+Nc_1'\,.
\ee
A similar expression holds in the case of a boundary, with half of the
$w_i$ corresponding to the image points (this is relevant in considering
the time evolution,\cite{cc-05} where the boundary, in the time axis, is 
the initial condition).

\section{Non-critical 1+1-dimensional models}
\label{noncrit}

So far, we considered the case of a system that is exactly at the 
quantum critical point. The natural question arising is what happens 
when a system is close to the phase transition, in the region where the 
correlation length $\xi$ is large, but finite. 
In this case the system is 
still effectively described by a QFT, that is massive, the mass being the
inverse of the correlation length.

When the subset $A$ is the whole negative real axis (and $B$ the positive 
one), we showed that the entanglement entropy is $S_A=(c/6) \log\xi/a$.
The argument we gave\cite{cc-04} is quite technical and we not 
repeat it here. 
However with this formula at hand, we can easily generalize it to more 
complicated situations, i.e., when $A$ consists of several disjoint 
intervals, having ${\cal A}$ boundary points with its complement. 
In fact, when the correlation length is large, but smaller than the length 
of each interval, the disjoint pieces must not interact (this is nothing but 
cluster decomposition in ordinary QFT), resulting in
\be
S_A={\cal A} \frac{c}{6}\log\frac{\xi}{a}\,.
\label{Snoncrit}
\ee 
When the correlation length exceeds the lengths of all the intervals we 
crosses over to the CFT results. In the intermediate regime (i.e. $\xi$ of 
the order of some interval length), we expect to see very complicated 
crossover scaling forms, that are still universal, but probably depending 
on the model in a very complicated way, not only through the central charge.
We emphasize once again that this result is true only when $\xi\gg a$.

\begin{figure}[t]
\centerline{\epsfig{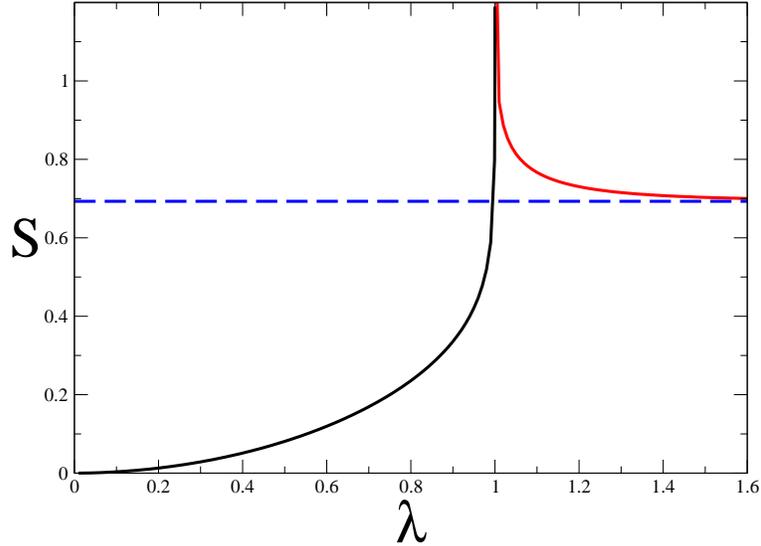}}
\caption{Entanglement entropy for the 1D Ising chain as function of $\lambda$.
}
\label{Fig}
\end{figure}

Equation (\ref{Snoncrit}) is completely general, and, at the time of 
publication of our work, there was no evidence for its validity (apart
from numerical results\cite{Vidal} suggesting the saturation of $S_A$ for 
finite $\xi$). For this reason we checked this formula in 
several models in the geometry with ${\cal A}=1$. 
We considered the massive Gaussian QFT, reproducing Eq.
(\ref{Snoncrit}) with $c=1$. 
In this particular case we also calculated\cite{cc-04} the entanglement 
entropy in a finite system of total length $2L$ divided into two 
equal parts, showing explicitly the crossover between the CFT and the massive
regimes.

Then, relaxing even the condition $\xi\gg a$, we 
solved exactly the Quantum Ising chain in a transverse field (by means of
Corner Transfer Matrix method\cite{pkl-99}) with hamiltonian
\be
H_I=-\sum_{n}\sigma^x_n-
\lambda  \sum_{n}\sigma^z_n\sigma^z_{n+1}\,,
\label{HamI}
\ee
that displays a quantum phase transition for $\lambda=1$. 
The final result we obtained is given by complicated (but rapidly converging)
infinite sums.\cite{cc-04} 
The resulting entanglement entropy as function of $\lambda$ is shown 
in Figure \ref{Fig}, in particular it diverges at $\lambda=1$, as it should. 
For $\xi=|\lambda-1|^{-1}\gg 1$, it behaves as 
$S_A\sim 1/12 \log\xi$, in agreement with Eq. (\ref{Snoncrit}) with $c=1/2$.
We also considered the XXZ model,\cite{cc-04} in the vicinity of the 
isotropic point, confirming Eq. (\ref{Snoncrit}) for $c=1$.

Later, Peschel\cite{p-05} was able to write the infinite sums in terms of 
elliptic integrals and also to generalize the result to the $XY$ chain 
in a transverse field. 
Then a direct comparison with the explicit formula for ${\cal A}=2$ (obtained
in the meantime by Its et al.\cite{ijk-04} using a completely different
approach) was possible, showing also the validity of the 
cluster decomposition.

\section*{Acknowledgments}

This work was supported in part by the EPSRC under Grant GR/R83712/01.

\end{document}